\documentclass[11pt]{article}
\newcommand{\rev}[1]{ {\color{black}{#1}}}
\baselineskip
\usepackage{amssymb}
\usepackage{graphicx}
\usepackage{latexsym}
\usepackage{epsfig}
\usepackage{amsmath}
\usepackage{color}
\usepackage[english]{babel}
\usepackage{comment}
\usepackage{caption}
\def\ii{\mathrm{i}}
\newcommand{\bra}{\left \langle}
\newcommand{\ket}{\right \rangle}
\newcommand{\rme}{\mathrm{e}}
\newcommand{\eq}{\begin{equation}}
\newcommand{\en}{\end{equation}}
\newcommand{\eqa}{\begin{eqnarray}}
\newcommand{\ena}{\end{eqnarray}}

\newcommand{\R}{\mathbb{R}}
\newcommand{\SO}{\mathrm{SO}}

\newcommand{\SNG}{S_{\mbox{\tiny{NG}}}}
\newcommand{\Scl}{S_{\mbox{\tiny{cl}}}}

\newcommand{\Z}{\bf{Z}}
\newcommand{\br}{\langle}
\newcommand{\kt}{\rangle}
\newcommand{\SU}{\mathrm{SU}}

\newcommand{\redchisq}{\chi^2_{\tiny\mbox{red}}}

\newcommand{\de}{\partial}
\newcommand{\tmb}[1]{{\mbox{\tiny{#1}}}}
\def\half{\frac{1}{2}}

\def\sumn{\sum_{n=1}^{\infty}}
\def\vev#1{\langle      #1 \rangle}

\begin{document}

\begin{titlepage}
\begin{center}
{\Large\bf Effective string description of the confining flux tube at finite temperature}
\end{center}
\vskip1.3cm
\centerline{Michele~Caselle}
\centerline{\sl Department of Physics, University of Turin \& INFN, Turin}
\centerline{\sl Via Pietro Giuria 1, I-10125 Turin, Italy}
\vskip1.0cm
\begin{abstract}
In this review,  after a general introduction to the
effective string theory (EST) description of confinement in pure gauge
theories,  we discuss the behaviour of EST as the
temperature is increased.
  We show that, as the deconfinement point is approached from below,
several universal features of confining gauge theories, like the ratio
  $T_c/\sqrt{\sigma_0}$, the linear increase of the squared width of the
  flux tube with the interquark distance, or the temperature dependence of the interquark potential, 
  can be accurately  predicted by the effective string.
Moreover in the vicinity of the deconfinement point  the   
EST behaviour
turns out to be in good agreement with what predicted by conformal
invariance or by dimensional reduction, thus further supporting the 
validity of an EST approach to confinement.

\end{abstract}

\end{titlepage}

\tableofcontents

\section{Introduction}
\label{sec:introduction}

One of the most powerful tools we have for studying the non-perturbative behaviour of confining Yang-Mills theories is the so called "Effective String Theory" (EST) in which the confining flux tube joining together a quark-antiquark pair is modeled as a thin vibrating 
string~\cite{Nambu:1974zg,Goto:1971ce,Luscher:1980ac,Luscher:1980fr,Polchinski:1991ax} .  
As explicitly stated in its definition, this model is only an effective large distance description of the flux tube and not an exact non-perturbative solution of the Yang-Mills theory, however, due to the peculiar features of the string action, it turns out to be a highly predictive effective model, whose results can be successfully compared with the most precise existing Monte Carlo simulations in Lattice Gauge Theories (LGTs).  
Besides its predictive power EST is also interesting from a theoretical point of view, since it is a perfect laboratory to test more refined nonperturbative descriptions of Yang-Mills theories, guess new hypothesis, and drive our understanding of the role of string theory in this game.

EST also plays an important role from a phenomenological point of view since it can be used to model the glueball spectrum of QCD 
\cite{Isgur:1984bm} or to model the large distance (non-perturbative) part of the interquark potential in heavy quarkonia~\cite{Luscher:1980ac,Luscher:1980fr} or the onset of the deconfinement transition \cite{Olesen:1985ej}.

The simplest, Lorentz invariant, EST is the Nambu-Goto model~\cite{Nambu:1974zg,Goto:1971ce} which will be the main subject of this review.  The Nambu-Goto action can be considered in this framework as a first order approximation of the actual EST describing the non-perturbative behavior of the Yang-Mills theory. 
The most interesting result of the last ten years of EST studies 
is that this first order approximation works remarkably well and agrees within the errors, \rev{in the large distance limit}, with almost all the existing Monte Carlo simulations for all the confining models that have been studied (with only few exceptions  \cite{Caselle:2014eka,Caselle:2016mqu}).
We shall see below that this is not by chance and that it is instead a direct consequence of the peculiar nature of the EST and of the strong constraining power of Lorentz invariance in this context.

This also explains the impressive universality of the infrared regime of confining gauge theories (with only a mild dependence on the number of space-time dimensions, exactly as predicted by the Nambu-Goto model), which show essentially the same behaviour for the interquark potential, the deconfinement temperature and the glueball spectrum.
This universality of LGT results, which holds for models as different as the three dimensional gauge Ising model and the four dimensional SU(3)  Yang-Mills theory was in the past one of the major puzzles in the Lattice community and is now understood only as a side effect of the impressive effectiveness of the Nambu-Goto approximation.  It is thus only an apparent universality and all the details on the gauge group are expected to be encoded in the higher order EST corrections beyond Nambu-Goto.
It is thus clear why lot of efforts have been devoted in these last years to the identification and modellization of these 
non-universal higher order corrections. The hope is that, since they depend on the particular type of confining gauge theory, they could  shed some light on specific non-perturbative properties of the theory, for instance on the
 non-pertubative degrees of freedom (say, instatons or monopoles) driving confinement in the model.
 
Due to the asymptotic large-distance nature of the EST expansion, the optimal regime in which one can observe higher order terms is for short interquark separations, much smaller than the length of the Polyakov loops. In the string language this is known as the "open string channel". 

While the above choice is the one which is more often used, it is easy to see, looking at the explicit expression of the EST partition function   
that the opposite choice (the "closed string channel" in string language) in which the short direction is the one with periodic boundary conditions, is better suited to observe  higher order corrections. In the language of Lattice Gauge Theories  this is the high temperature regime of the theory  in which the temperature is just below the deconfinement transition. This is exactly the limit in which we are interested in this review. 

While there are already several good general reviews on EST (see for instance \cite{ Bali:2000gf,Aharony:2013ipa,Brandt:2016xsp}), the goal of this paper is to focus specifically on the EST behaviour in the high-$T$ regime and to discuss how our understanding of Lattice Gauge Theory in this limit can help us to constrain and check EST. 
In particular an important reason of interest of this limit is that, in LGTs with a second order deconfinement transition, several non trivial results on EST can be obtained using renormalization group arguments of the type discussed in~\cite{Svetitsky:1982gs}. This approach, which allows to map a $(d+1)$ dimensional LGT into a suitably chosen $d$ dimensional spin model,  will be one of the main focus of the present review.

In this review we shall mainly focus on two observables: the interquark potential and the flux tube width which in the past years played a major role in the progress of our understanding of EST properties. In particular, the review is organized as follows.  
Sect.2 will be devoted to a brief introduction to Lattice Gauge Theories. In Sect.3 we shall discuss the main properties of
EST (and in particular of the Nambu-Goto action) with a particular focus on the high-$T$ behaviour. Then in sect.4 we shall compare EST predictions for the interquark potential in the high temperature regime with Monte Carlo simulations.  In sect.5 we shall address the important issue of the width of the flux tube and discuss its behaviour in the high-$T$ regime. Sect.6 will be devoted to a few concluding remarks.

  \section{A brief summary of LGTs.}
  \label{sect:LGT}
  The natural context in which we can see the EST at work is in the confining phase of Lattice Gauge Theories (LGT) where EST is expected to describe the large distance behaviour of the confining flux tube joining a quark antiquark pair. Thus in order to fix notations and to better understand the physics behind EST it is useful to briefly discuss a few basic notions of LGTs.  
We refer the interested reader to the book \cite{Montvay:1994cy} for a more detailed introduction to LGTs.
  
\vskip.3 cm
The partition function of a gauge theory in $D$ spacetime dimensions
 with gauge group $G$ 
regularized on a lattice is
\eq
Z=\int\prod \mbox{d}U_\mu(\vec x,t) 
\exp\{-\beta\sum_p\, \mbox{Re}\, \mbox{Tr}(1-U_p)\}~~~,
\label{eq1}
\en
where $U_\mu(\vec x,t)\in G$ is the link variable
at the site  $(\vec x,t)=(x_1,..,x_{D-1},t)$ in the direction $\mu$ and
$U_p$ is the product of the 
links around the plaquette $p$.

We shall denote in the following with $N_t$ ($N_s$) the lattice size in the
time (space) direction and assume for simplicity $N_s$ to be 
the same for all the space-like directions. We shall use $d$ to denote the number of space-like directions in the lattice, thus $D=d+1$.
To simplify notations we shall
fix the lattice spacing $a$ to 1 and neglect it in the following.

As it is well known the link variable $U_\mu(\vec x,t)$ is {\sl not} gauge invariant and only
the traces of ordered products of link variables along closed paths are gauge invariant.

The simplest choice is the Wilson loops
\eq
W(\gamma)~=~{\rm Tr} \prod_{(\vec x,t)\in\gamma}U_\mu(\vec x,t)
\en
where the product is assumed to be ordered along the path $\gamma$. If we choose the path $\gamma$
 to be a rectangle of size $R\times L$ (with $L$  along the Euclidean "time" direction and $R$ along one of the space directions) then it is possible to relate the expectation value of $W(R\times L)$ to the interquark potential as:
\eq
V(R)~=~-\lim_{L\to\infty} \frac{1}{L} \log\br W(R,L)\kt~~~.
\label{lu0}
\en

The idea behind this definition is that we may think of $\vev{W(R,L)}$ 
as the free energy due to the
creation at the time $t_0$ of a 
quark and an antiquark pair  which are instantaneously
moved at a distance $R$ from each other, keep their position
for a time $L$ and finally annihilate at the instant $t_0+L$. 

A confining LGT will be characterized by  a linearly rising potential and thus, according to
eq.(\ref{lu0}) we expect for the Wilson loop an "area law" of this type

\eq
\br W(R,T)\kt ~\sim~ e^{-\sigma_0 RT+p(R+T)+k}~.
\label{area}
\en
 The area term is responsible for confinement 
while the perimeter and constant terms are non universal contributions related
to the discretization procedure.
The physically important quantity is the coefficient of the
 area term which represents the lattice estimate 
  of the string tension.

\subsection{Finite temperature LGTs}
\label{s2.1}

It is important at this point to stress that eq.(\ref{lu0}) above, only defines the so called {\sl zero temperature} interquark potential and accordingly $\sigma_0$ is the {\sl zero temperature} string tension. If one is interested in the finite temperature behaviour of the interquark potential and in the possible presence of a deconfinement transition at some finite temperature $T_c$ 
the lattice regularization prescription must be modified.

The lattice regularization of a generic Quantum Field Theory (QFT) at a non-zero, finite temperature $T$ can be obtained by \rev{imposing periodic boundary conditions in the time direction for the bosonic field (and antiperiodic for fermionic ones)}. With this choice
 a lattice of size $(N_sa)^{d}(N_ta)$  represents the regularized version of 
a system of finite volume $V=(N_sa)^{d}$  at a finite temperature
$T=1/N_ta$\footnote{\rev{Even if in the rest of the paper, having set $a=1$, we shall systematically  use  $N_t$ and $N_s$ as 
a shorthand notation for $N_ta$ and $N_sa$, in the above formulas we restored the lattice spacing to emphasize the correct dimensions of the physical quantities we are defining.}}. The compactified "time" direction at this point does not have any longer the meaning of time (recall that we are describing a system at equilibrium in the canonical ensemble) but its size $N_ta$ is instead a measure of the inverse temperature of the system $N_ta=1/T$.
 
As a consequence, in a finite temperature setting, the Wilson loop cannot be related any more to the interquark potential as we did above.
Fortunately we have a different way to construct a quantity with the same physical interpretation.
In a finite temperature setting one can define
a new class of gauge invariant observables which are usually called
 Polyakov loops.
 
 A Polyakov loop $P(\vec x)$ is the trace of the ordered product of all
 time-like links with the same space-like coordinates; this loop is
 closed owing  to the periodic boundary conditions in the time direction:
\eq
P(\vec x)=\mbox{Tr}\prod_{z=1}^{N_t}U_t(\vec x,z)~~~.
\en

In a pure LGT  the Polyakov loop has 
a deep physical meaning, since its  expectation value is related to the
free energy of a single
isolated quark. 
Hence the fact that the Polyakov loop acquires a non-zero expectation 
value can be considered as a signature of deconfinement and 
the Polyakov loop is  thus the
order parameter of such a deconfinement transition.

 The value $\beta_c(N_t)$ of this deconfinement 
transition in a lattice of size $N_t=1/T$ in the compactified time direction can be used to define a new 
physical observable $T_c$ which is obtained by inverting $\beta_c(N_t)$. We obtain in this way, for each value of $\beta$, 
the lattice size in the time direction (which we shall call in
the following $N_{t,c}(\beta)$) at which the model undergoes the deconfinement
transition and from this the critical temperature $T_c(\beta)\equiv 1/N_{t,c}(\beta)$ as a function of $\beta$
\vskip.3 cm

\subsection{The finite temperature interquark potential.}
In a  finite temperature setting the interquark potential can be extracted by looking at the correlations  
of Polyakov loops in the confined phase.
The correlation of two loops $P(x)$  at
a distance $R$ and at a temperature $T=1/N_t$ (which we denote with the subscript $N_t$ in the expectation value) 
is given by 

\eq
\langle P(x)P^\dagger(x+R) \rangle_{N_t} ~\equiv ~ {\rm e}^{-\frac{1}{T}V(R,T)}~ = ~{\rm e}^{- N_t V(R,T)}~~,
\label{polya}
\en
\noindent
where we consider the free energy $V(R,N_t)$ as a proxy for the interquark potential at a finite temperature $T$ 
\eq
V(R,T)=-\frac{1}{N_t}\log{\langle P(x)P^\dagger(x+R) \rangle_{N_t}  }~~~~.
\label{potft}
\en

\noindent
If we assume also for this correlator an area law similar to the one discussed above for the Wilson loop:

\eq
\langle P(x)P^\dagger(x+R) \rangle_{N_t}  \sim {\rm e}^{-\sigma(T) N_t R}
~~~,\label{area2}
\en
then we find again a confining behaviour for the interquark potential. In the above equation 
 $\sigma(T)$ denotes this time the {\sl finite temperature} string tension. As we shall see below $\sigma(T)$ is a decreasing function of $T$ and vanishes exactly at the deconfinement point \cite{Kaczmarek:1999mm,Cardoso:2011hh}. Following the definitions of the previous section, we may identify $V(R)$ with the $T\to 0$ limit of $V(R,T)$ and $\sigma_0$ with the 
$T\to 0$ limit of $\sigma(T)$.

It is interesting to notice that
the observable (\ref{polya}) is similar to the expectation value of an ordinary
 Wilson loop except for the boundary conditions, which are  in this case 
fixed in the space directions and periodic in the time direction. The
 resulting geometry is that of a cylinder, which is topologically 
different from the rectangular geometry of the Wilson loop.

\subsection{Center symmetry and the Polyakov loop}
\label{centersec}
The major consequence of the periodic boundary conditions in the time direction
is the appearance of a new global symmetry of the action, with symmetry  
group the center $C$ of the gauge group (i.e. ${\bf Z}_N$ if the gauge group is $SU(N)$).

This symmetry can be realized, for instance, by acting on 
all the timelike links 
of a given space-like slice with the same element $W_0$ belonging to the center
of the gauge group.
\eq
U_t(\vec{x},t) \to ~W_0~U_t(\vec{x},t)  \hskip 1cm \forall\, \vec{x},~~~t~
\mbox{fixed}~.
\en
it is easy to see that the Wilson action is invariant under such transformation. 
while the Polyakov loop  transforms as:
\eq
P(\vec{ x}) \to~ W_0~ P(\vec{ x})~;
\label{polya3}
\en
 thus it is a natural order parameter for this symmetry.
It will acquire a non zero expectation value if the center symmetry is
spontaneously broken. 

\rev{Thus we see that the Polyakov loop is at the same time the order paramater of the deconfinement transition and of the center symmetry: the deconfinement transition in a pure lattice gauge theory coincides with the center
symmetry breaking phase transition.}  In the deconfined phase the center symmetry is spontaneously broken while in the confining
 phase it is conserved.

\subsection{The Svetitsky--Yaffe conjecture}
\label{sysec}
The peculiar role played by the Polyakov loops in the above discussion,
suggests to use some kind  of effective action for the Polyakov loops, \rev{integrating out the spacelike links of the model},
to study the deconfinement
transition and, more generally, the physics of finite temperature LGT. Such a construction corresponds in all respects to a "dimensional reduction":
starting from a ($d+1$) dimensional LGT we end up with an effective action for the Polyakov loops which will be a
$d$ dimensional spin model with global symmetry the center of the original gauge group.

While the explicit construction of such an effective action may be cumbersome and can be performed only as a strong coupling expansion,
some general insight on the behaviour of the model can be deduced by simple renormalization group arguments~\cite{Svetitsky:1982gs} .

Indeed, even if as a result of the integration over the original gauge degrees of freedom we may expect long range interactions between the Polyakov loops, 
it can be shown~\cite{Svetitsky:1982gs} that these interactions decrease exponentially with the distance. 
Thus, if the phase transition is continuous, in the vicinity of the critical point the fine details of the interactions can be neglected, and the model will belong to the same universality class of the simplest spin model, with only nearest neighbour interactions, sharing the same symmetry breaking pattern.  
For instance, the deconfinement transition of the $SU(2)$ LGT in (2+1) and (3+1) dimensions, which in both cases is continuous, belong to the same universality class as, respectively, the two dimensional  and the three dimensional Ising models.

This mapping has several important consequences:
\begin{itemize}
\item[{\em a)}]
The ordered (low temperature) phase of the spin model corresponds to the 
deconfined (high temperature) phase of the original gauge theory. This is the phase in which both the Polyakov loop, in the original LGT, and the spin, in the effective spin model, acquire a non-zero expectation value.  
\item[{\em b)}]
As for the operator content of the two models,  the Polyakov loop is mapped into the spin operator, while the plaquette is mapped into the energy operator of the effective spin model. Accordingly, the Polyakov loop correlator in the confining phase,  from which we extract the interquark potential, is mapped into the spin-spin correlator of the disordered, high temperature phase of the spin model

\item[{\em c)}]
Thermal perturbations from the critical point in the original gauge theory, which are driven by the plaquette operator, are mapped into thermal perturbation of the effective spin model which are driven by the energy operator. Notice however the change in sign: an {\sl increase} in temperature of the original gauge theory corresponds to a {\sl decrease} of the temperature of the effective spin model. 

\end{itemize}

A major consequence of this correspondence is that, in the vicinity of the deconfinement point, the behaviour of the interquark potential is strongly constrained and thus it represents, as we shall see, a powerful tool to test the predictions of the effective string model.

\subsection{EST versus LGT: the roughening transition}
As we mentioned above, a confining interquark potential implies an area law for the Wilson loop (at zero temperature) or for the correlator of two Polyakov loops (at finite temperature).  A nice feature of the lattice regularization is that such an area law naturally arises from a strong coupling expansion of these observables. Order by order in the strong coupling parameter $\beta$, the expectation value of a Wilson loop (or of a Polyakov loops correlator) is described by the sum over all the possible surfaces 
bordered by the Wilson loop with a weight proportional to their area. As it is well known this expansion diverges at the so called "roughening point"\cite{Luscher:1980ac,Luscher:1980fr,Hasenfratz:1980ue,Itzykson:1980fz}, well before the values of $\beta$ for which a continuum limit of the lattice regularization can be approached. 
 This roughening transition is due to the vanishing of the stiffness of the strong coupling surfaces and has a very insightful explanation from an EST point of view. 
The vanishing of the surface stiffness ensures that the surfaces bordered by the Wilson loop can freely fluctuate as actual continuum-like 
surfaces and that they are not any more anchored to the crystallographic planes of the lattice and can thus be described by a set of $(D-2)$ 
real degrees of freedom representing their transverse displacement from the Wilson loop plane \cite{Hasenfratz:1980ue,Itzykson:1980fz}.  Upon quantization these transverse coordinates will become the $(D-2)$ bosonic degrees of freedom  of the EST description which we shall discuss in the next section\cite{Luscher:1980ac,Luscher:1980fr}.  These massless quantum fluctuations delocalize the flux tube
 which acquires a nonzero width, which diverges logarithmically as the
 interquark distance increases~\cite{Luscher:1980iy}. We shall discuss in detail this issue in sect.\ref{sect:width}.

We may summarize all these observations by saying that the LGT regularization strongly supports an Effective String Theory description of confinement.
We shall devote the next section to a precise formulation of this EST.

\section{Effective String description of the interquark potential.}
\label{sec:est}
Even if a rigorous proof of quark confinement in Yang-Mills theories is still missing, 
there is little doubt that confinement  is associated to the formation of a thin string-like 
flux tube~\cite{Nambu:1974zg,Goto:1971ce,Luscher:1980ac,Luscher:1980fr,Polchinski:1991ax},
 which generates, for large quark separations, a linearly  rising  confining potential.

This picture is strongly supported by the lattice regularization of Yang-Mills theories where, as we have seen in the previous section, the vacuum expectation value of Polyakov loops correlators is given by a sum over certain lattice surfaces which can be considered as the 
world-sheet of the underlying confining string.

This picture led  L\"uscher and collaborators ~\cite{Luscher:1980ac,Luscher:1980fr}, more than forty years ago, to propose that 
 the dynamics of the flux tube for large interquark distances could be described by a free massless bosonic field theory in two dimensions.  
\eq
S[X]=S_{cl}+S_0[X]+\dots,
\label{frees}
\en
where the classical action $S_{cl}$ describes the usual perimeter-area term,
$X$ denotes the two-dimensional bosonic 
fields $X_i(\xi_1,\xi_2)$, with  $i=1,2,\dots, D-2$,     
describing the 
transverse displacements of the string with respect the configuration 
of minimal energy, $\xi_1,\xi_2$ are the coordinates on the world-sheet
and $S_0[X]$ is the Gaussian action
\eq
S_0[X]=\frac{\sigma_0}{2}\int d^2\xi\left(\partial_\alpha X\cdot\partial^\alpha X
\right) ~
\label{gauss}
\en
We are assuming an Euclidean signature for both the worldsheet and the target space.
 
This is the first example of an effective string action and, as we shall see below, it is actually nothing else than the large distance limit of the Nambu-Goto string written in the so called "physical gauge".
 
This free Gaussian action can be easily integrated, leading in the $T\to 0$  ($N_t\to \infty$) limit to a correction to the linear quark-anti-quark potential, known as  L\"uscher term \cite{{Luscher:1980ac},{Luscher:1980fr}}
\eq
V(R)=\sigma_0 R  + c -\frac{\pi(D-2)}{24R}+O(1/R^2)~.
\label{freeV}
\en
We shall neglect from now on the constant $c$ which is related to the perimeter term discussed in the previous section.

It is instructive to look at this correction for finite values of $N_t$.  Thanks to the Gaussian nature of the action the integration can be easily performed also for finite values of $N_t$, for instance using the $\zeta$ function regularization, leading to the following 
result~\cite{Dietz:1982uc,Alvarez:1981kc, Arvis1983b}:

\eq
V(R,T)=\sigma_0 R~+~ \frac{D-2}{N_t}\log\left({\eta(q)}\right)~~~,
\label{bos}
\en
\noindent
where $\eta$ denotes the Dedekind eta function (see the Appendix \ref{AppendixA}):
\eq
\eta(\tau)=q^\frac{1}{24}\prod_{n=1}^\infty(1-q^n)\hskip0.5cm
;\hskip0.5cmq=e^{2\pi i\tau}
\hskip0.5cm
;\hskip0.5cm \tau=i\frac{N_t}{2R}~~~.
\label{eta}
\en
\vskip.3cm

To understand the meaning of this result it is useful to expand it in the two limits $R\ll N_t$ and $R\gg N_t$.

\begin{description}
\item{$R\ll N_t$,~~~  low temperature}
\eq
V(R,T)=\sigma_0 R + \left[-\frac{\pi }{24 R}
+\frac{1}{N_t}\sum_{n=1}^\infty \log (1-e^{-\pi nN_t/R})\right](D-2)~~~,
\label{eta1}
\en

\item{$R\gg N_t$,~~~   high temperature}   
\eq
V(R,T)=\sigma_0 R + \frac{D-2}{N_t}\left[-\frac{\pi R}{6 N_t}+\frac{1}{2} \log\frac{2R}{N_t}
+\sum_{n=1}^\infty \log (1-e^{-4\pi nR/N_t})\right]~~~.
\label{eta2}
\en
\end{description}

 From a string point of view these limits correspond to the open and closed  string channels respectively. They are related by a modular transformation
$\tau\to -1/\tau$ 
\begin{align}
\label{Seta}
\eta\left(\rme^{- 2\pi\ii/\tau}\right) = \sqrt{-\ii\tau} \eta\left(\rme^{2\pi\ii\tau}\right)~.
\end{align}

which is known as open-closed string duality .

In the LGT language the two limits correspond respectively to the low temperature and the high temperature limits where, obviously,  with high temperature we mean a value of $T$ large,  but still below the deconfinement temperature, so that  a confining flux tube still exists between the quark and the antiquark and an EST picture is still a valid description of the infrared behaviour of the theory.

It is interesting to see that the EST corrections have a completely different behaviour in the two regimes.

At low temperature we find a rather mild correction, which is dominated by the L\"uscher term mentioned above (the first term in eq.(\ref{eta1})) while the remaining terms vanish in the $N_t\to \infty$ limit.

On the contrary at high temperature we find that the dominant term is linear in $R$ and gives a large correction,  which increases as the temperature increases, and counteracts the string tension. 
\eq
V(R,T)\sim \left(\sigma_0 - \frac{\pi (D-2)}{6 N_t^2}\right)~R~~~~.
\en
We shall see below that, if one studies the whole Nambu-Goto action this correction represents 
 only the first term of an infinite set of corrections which can be resummed as follows

\eq
V(R,T)~\sim~ \sigma_0 \sqrt{1- \frac{\pi (D-2)}{3 \sigma_0 N_t^2}}~R~ \equiv~ \sigma(T) R
\en
where we have introduced a temperature dependent string tension $\sigma(T)$ defined as:

\eq
\sigma(T)= \sigma_0 \sqrt{1- \frac{\pi (D-2)}{3 \sigma_0 N_t^2}}
\label{sigma_T}
\en
Intuitively, what is happening in this regime is that the fluctuations induced by the temperature tend to reduce the confining force of the flux tube.  As the temperature increases, fluctuations get stronger and stronger and finally, at the deconfinement point, the flux tube is destroyed by the fluctuations and there is no more a confining potential between the quark and the antiquark.

It is exactly this finite temperature regime the main focus of the present review, and it is clear now the reason of this choice:  in this regime  string effects are magnified and can be more easily compared with numerical simulations.

\subsection{The Nambu-Goto action}

It is easy to see that the free Gaussian action discussed above cannot be a consistent effective string description of the flux tube since it does not fulfill the constraints imposed by the  Lorentz invariance of the original gauge theory (we shall discuss this issue in more detail below). The simplest possible EST fulfilling these constraints is the well known Nambu-Goto action \cite{Nambu:1974zg,Goto:1971ce}. 
As we shall see below the free Gaussian action of eq.(\ref{gauss}) is actually the first term of the large distance expansion of the NG action. This explains why, notwithstanding its lack of consistency, its predictions, and in particular the L\"uscher term, were initially found in good agreement with LGT simulations of several different gauge models \cite{{Hasenbusch:1992zz},{Caselle:1996ii},Ambjorn:1984me,Ambjorn:1984yu,Necco:2001xg,
{Juge:2002br},{Lucini:2001nv},{Luscher:2002qv}}, and why, with the improvement of LGT simulations, this agreement was later shown to hold only for for Polyakov loops correlators with large separations and higher order corrections (in particular the next to leading Nambu-Goto term that we shall discuss below) started to be detected \cite{{Caselle:1994df},{Lucini:2002wg},{Caselle:2004er},{Caselle:2005xy},{Bringoltz:2006zg},{Caselle:2006dv},{HariDass:2006pq},{Bringoltz:2008nd},{Athenodorou:2011rx},
{Athenodorou:2010cs},{Caselle:2010pf},{Billo:2011fd},{Mykkanen:2012dv},{Athenodorou:2013ioa},{Caselle:2016wsw},{Athenodorou:2016kpd}}\footnote{Notice that some of the first studies on EST were actually performed in the three dimensional Ising model. In these simulations instead of the interquark potential one studies the interface free energy which is also described by the EST, but with different boundary conditions. In particular this is the case of the following papers:
\cite{{Hasenbusch:1992zz},Caselle:1994df,{Caselle:2006dv},Caselle:2016wsw}.}.

In the Nambu-Goto model~\cite{Nambu:1974zg,Goto:1971ce}, the string action 
$S_\tmb{NG}$ is :
\begin{align}
\label{NGaction}
S_\tmb{NG}= \sigma_0 \int_\Sigma d^2\xi \sqrt{g}~,
\end{align} 
where $~g\equiv \det g_{\alpha\beta}~$ and
\begin{align}
\label{NGaction2}
g_{\alpha\beta}=\partial_\alpha X_\mu~\partial_\beta X^\mu
\end{align} 

is the induced metric on the reference world-sheet surface $\Sigma$ and, as above, we denote the worldsheet coordinates as $\xi\equiv(\xi^0,\xi^1)$. This term has a simple geometric interpretation: it measures the area of the surface spanned by the string in the target space and is thus the natural EST realization of the sum over surfaces weighted by their area in the rough phase of the LGT model which we discussed above. This action has only one free parameter:  the string tension $\sigma_0$ which has dimension $(\mbox{length})^{-2}$.  Once this is fixed, say,  by a fit to the large distance behaviour of the lattice data at zero temperature, there are no more free degrees of freedom in the model which is thus, as we shall see,  highly predictive.

In order to perform calculations with the Nambu-Goto action one has first to fix its reparametrization invariance.  The standard choice is the so called ``physical gauge''. In this gauge the two worldsheet coordinates are identified with the longitudinal degrees of freedom of the string: $\xi^0=X^0$, $\xi^1=X^1$, so that the string action can be expressed as a function only of the $(D-2)$ degrees of freedom corresponding to the transverse displacements, $X^i$, with $i=2, \dots , (D-1)$ which are assumed to be single-valued functions of the worldsheet coordinates. We shall comment below on the problems of this gauge fixing choice, but let us assume it for the moment and let us see what are the consequences.

With this gauge choice  the determinant of the metric can be written as
\eqa
g&=&1+\de_0 X_i\de_0 X^i+\de_1 X_i\de_1 X^i\nonumber\\
&&\ \ \ +\de_0 X_i\de_0 X^i\de_1 X_j\de_1 X^j
-(\de_0 X_i\de_1 X^i)^2
\ena

and
the Nambu-Goto action can then be written as a low-energy expansion in the number of derivatives of the transverse degrees of freedom of the string which, by a suitable redefinition of the fields, can be rephrased as a large distance expansion. The first few terms in this expansion are
 
 \begin{equation}
S=\Scl+\frac{\sigma_0}{2}\int d^2\xi\left[\partial_\alpha X_i\cdot\partial^\alpha X^i+
\frac18(\partial_\alpha X_i \cdot\partial^\alpha X^i)^2
-\frac14(\partial_\alpha X_i \cdot\partial_\beta X^i)^2+\dots\right],
\label{action2NG}
\end{equation}

and we see, as anticipated, that the first term of the expansion is exactly the Gaussian action of eq.(\ref{gauss}).
From a Quantum Field Theory point of view the free Gaussian action is the two dimensional
 Conformal Field Theory (CFT) of the $D-2$ free bosons which represent the transverse degrees of freedom.

Remarkably enough, it can be shown that all the additional terms in the expansion beyond the Gaussian one combine themselves so as to give 
 an exactly integrable, irrelevant perturbation of the Gaussian term \cite{Dubovsky:2012sh}, driven by the $T\bar T$ operator of the $D-2$ free bosons \cite{Caselle:2013dra}.

Thanks to this exact integrability, the partition function of the model can be written explicitely\footnote{The explicit expression for the partition function was actually found even before this $T\bar T$ study,  first by using the constraints imposed by the open-closed string duality~\cite{Luscher:2004ib} and then using a d-brane formalism \cite{Billo:2005iv}.}\cite{Aharony:2018bad,Datta:2018thy}. 
For the Polyakov loop correlator in which we are interested here \footnote{Similar expressions can be obtained also for the other relevant geometries: the Wilson loop \cite{Billo:2011fd} and the interface \cite{Billo:2006zg}}, the expression in $D$ space-time dimensions is, using the notations of  \cite{Luscher:2004ib, Billo:2005iv}:
\begin{equation}
  \left\langle P(x)^{\ast}\kern-1pt P(y)\right\rangle
  =\sum_{n=0}^{\infty}w_n\frac{2R \sigma_0 N_t}{{E}_n}
   \left(\frac{\pi}{\sigma_0}\right)^{\frac{1}{2}(D-2)}
  \left(\frac{{E}_n}{2\pi R}\right)^{\frac{1}{2}(D-1)}
  K_{\frac{1}{2}(D-3)}({E}_nR)
\label{NG}
\end{equation}
where $R$ denotes, as above, the interquark distance $R=|x-y|$, $w_n$ the multiplicity of the closed string states which propagate from one Polyakov loop to the other, and $E_n$ their energies which are given by
\begin{equation}
  {E}_n=\sigma_0 N_t
  \sqrt{1+\frac{8\pi}{\sigma_0 N_t^2}\left[-\frac{1}{24}\left(D-2\right)+n\right]}.
\label{energylevels}
\end{equation}

At large distance the correlator is dominated by the lowest state 
\eq
E_0=\sigma_0 N_t
  \sqrt{1-\frac{\pi (D-2)}{3\sigma_0 N_t^2}} = \sigma(T) N_t .
\label{E0}
\en  
  where $\sigma(T)$ is the finite temperature string tension defined in eq.(\ref{sigma_T}).

The weights $w_n$ can be easily obtained from the expansion in series of $q$ of the infinite products contained in the Dedekind functions which describes the large-$R$ limit of eq.~(\ref{NG}) (see ref.~\cite{Billo:2005iv} for a detailed derivation):
\begin{equation}
\label{etaexp}
\left(\prod_{r=1}^\infty \frac{1}{1 - q^r}\right)^{D-2}
= \sum_{k=0}^\infty w_k q^k.
\end{equation}
For $D=3$ we have simply $w_k=p_k$, the number of partitions of the integer $k$, while for $D>3$ these weights can be straightforwardly obtained from combinations of the $p_k$.

These weights diverge exponentially as $n$ increases; in particular we have:
\begin{equation}
w_n \sim \exp\left({\pi\sqrt{\frac{2(D-2)n}{3}}} \right)\;.
\label{eq26}
\end{equation}

Again, it is easy to see that the large distance expansion of eq.~(\ref{NG}) exactly matches the free Gaussian result of eq.(\ref{bos})

\subsection{The Nambu-Goto action at finite temperature.}

Looking at 
eq.~(\ref{NG}) we see that the NG partition function coincides with a
 a collection of free particles of mass ${E}_n$ and multiplicity $w_n$ in $D-1$ dimensions. 
 In the large distance limit only the lowest of these masses survives and the Polyakov loop correlator is described by an expression of this type
 \begin{equation}
  \left\langle P(x)^{\ast}\kern-1pt P(y)\right\rangle
  \sim
  \left(\frac{1}{R}\right)^{\frac{1}{2}(D-3)}
  K_{\frac{1}{2}(D-3)}({E}_0R)
\label{NG2}
\end{equation}
Remarkably enough this is exactly what we would expect from the Renormalization Group analysis of sect.\ref{sysec}.
In fact, if we interpret the Polyakov loop as a spin of a $D-1$ dimensional spin model with global symmetry the center of the gauge group, then, if the symmetry group is discrete (like for instance for the $SU(2)$ or $SU(3)$ LGTs), in the symmetric phase of the model the spin-spin correlator is described by an isolated pole in the Fourier space, which, when transformed back to the coordinate space becomes exactly the expression of eq.(\ref{NG2}).
In this interpretation, the mass $E_0$ becomes the inverse of the correlation length $\xi$ of the system.  We thus find (see eq.(\ref{E0}))

\begin{equation}
\frac{1}{\xi}=\sigma_0 N_t
  \sqrt{1-\frac{\pi(D-2)}{3 \sigma_0 N_t^2}
  }=\frac{\sigma_0}{T}  
  \sqrt{1-\frac{\pi(D-2)T^2}{3 \sigma_0}}
\label{xi_string}
\end{equation}

It is interesting to look at the large distance expansion of the interquark potential in this regime.  As anticipated the dominant term is linear in $R$ and is proportional to the finite temperature string tension $\sigma(T)$. On top of this we have a set of subleading corrections, (encoded in the asymptotic expansion of the modified Bessel function $K_{\frac{D-3}{2}}$) which represent a  specific signature of the Nambu-Goto action.

Using the large distance expansion of the modified Bessel function $K_n(z)$:
\eq
\label{K0_expansion}
K_n(z) = \sqrt{\frac{\pi}{2z}} e^{-z} \left[ 1 + \frac{4n^2-1}{8z} + \frac{16n^4-40n^2+9}{128z^2}+ \mathcal{O}(z^{-3})  \right]
\en
and the definition of the interquark potential in eq.(\ref{potft})
we find,  using eq.(\ref{NG2}) 
\eq
\label{potential_approximate_form}
V(R,N_t) \sim RT E_0  + \frac{T(D-2)}{2} \ln R + \frac{T(1-(D-3)^2)}{8 R E_0} ~\cdots
\en
where we dropped an irrelevant additive constant, and neglected terms which are suppressed by higher powers of $(RT)^{-1}$. 
In the following we shall mainly study models in $D=3$ dimensions.  In this case the above expression becomes:

\eq
\label{potential_D=3}
V(R,N_t) \sim R \sigma(T) + \frac{T}{2} \ln R + \frac{T^2}{8 R \sigma(T)}  ~\cdots
\en

In the framework of the Nambu-Goto approximation one can also derive an estimate of the critical temperature $T_{c,NG}$  measured in units of the square root of the string tension $\sqrt{\sigma_0}$ \cite{Olesen:1985ej,Olesen:1985ni,Pisarski:1982cn}
\begin{equation}
\frac{T_{c,NG}}{\sqrt{\sigma_0}}=\sqrt{\frac{3}{\pi(D-2)}}
\end{equation}
given by the value of the ratio $\frac{T_{c,NG}}{\sqrt{\sigma_0}}$ for which the lowest mass $E_0$ vanishes.
We can thus rewrite the energy levels as a function of $T/T_{c,NG}$ as
\begin{equation}
  {E}_n= \frac{(D-2)\pi T_{c,NG}^2}{3 T} \sqrt{ 1 - \, \frac{T^2}{T_{c,NG}^2}\left[1-\frac{24n}{D-2}\right]}.
\end{equation}

In this framework the correlation length can be written as:
\begin{equation}
  \xi(T)= \frac{3T}{(D-2)\pi T_{c,NG}^2} \frac{1}{\sqrt{ 1 - \frac{T^2}{T_{c,NG}^2} }}.
\end{equation}

which diverges as expected at the critical point.  This result is particularly interesting from a conceptual point of view since it makes explicit in which sense the Nambu-Goto action is an approximation of the "correct" effective string action. The critical index that we find: $\nu=1/2$ is the typical signature of the mean field approximation. We know from the 
Svetitsky-Yaffe analysis that this cannot be the correct answer and that the critical index should instead be that of the symmetry breaking phase transition of the $(D-1)$ dimensional spin model with symmetry group the center of the original gauge group. For instance,  for the $(3+1)$ dimensional $SU(2)$ model we expect to find 
$\nu=0.6299709(40)$~\cite{Komargodski:2016auf,Kos:2016ysd} which is the value for the three dimensional Ising model or for the $(2+1)$ $SU(2)$ LGT we expect $\nu=1$,  which is the cirtical index for the 2d Ising model. Besides this anomalous dimension, the major effect of the mean field approximation is, as usual, a shift in the critical temperature. Indeed, while the Nambu-Goto prediction for the deconfinement temperature is in four dimensions $T_{c,NG}=\sqrt{\frac{3\sigma_0}{2\pi}}\sim= 0.691 \sqrt{\sigma_0}$ the actual value for the $SU(2)$ deconfinement transition is slightly larger: $T_c/\sqrt{\sigma_0}= 0.7091(36)$ \cite{Lucini:2003zr}.
The fact that this shift is so small is another evidence of the goodness of the Nambu-Goto approximation. It is interesting to notice  that this agreement holds for all the LGTs which have been studied \cite{Lucini:2002ku,Lucini:2003zr,Lucini:2005vg,Liddle:2008kk,Lau:2015cna}, both in (2+1) and in (3+1) dimensions, with the only exception of the 3d $U(1)$ model \cite{Borisenko:2015jea}, for which, in fact, a different EST is expected \cite{Caselle:2014eka,Caselle:2019khe},  with a dominant contribution from the extrinsic curvature term.

\subsection{Beyond Nambu-Goto}
\label{beyond}
 We have seen from the above analysis that the Nambu-Goto action alone cannot be the end of the story. 
Finding hints of the correct EST action beyond the Nambu-Goto term is one of the major open challenges in this context.
We shall devote this subsection and the following to a brief discussion of this issue.

As a starting point let us notice that, from an effective action point of view, there is no reason to constrain 
the coefficients of the higher order terms in eq.(\ref{action2NG})  to the values
 displayed there. In principle, one should instead assume the most general form for such an effective action

 \begin{equation}
S=\Scl+\frac{\sigma_0}{2}\int d^2\xi\left[\partial_\alpha X_i\cdot\partial^\alpha X^i+
c_2(\partial_\alpha X_i \cdot\partial^\alpha X^i)^2
+c_3(\partial_\alpha X_i \cdot\partial_\beta X^i)^2+\dots\right],
\label{action2}
\end{equation}

and then fix the coefficients order by order using Monte Carlo simulations or experimental results.
However, one of the most interesting results of the last few years is that the $c_i$ coefficients are not 
arbitrary, but must satisfy a set of constraints to enforce the Poincarè invariance of the lattice gauge theory in the $D$ dimensional target space. These constraints were first obtained by comparing the string partition function in different channels,  using the open-closed string duality~\cite{Luscher:2004ib, Aharony:2009gg}. It was later realized~\cite{Meyer:2006qx, Aharony:2011gb, Gliozzi:2011hj, Gliozzi:2012cx, Meineri:2013ew} that they could be directly obtained  as a consequence of the Poincar\'e symmetry of the underlying Yang-Mills theory. \rev{A similar result, for the first few coefficients of the EST, was obtained also in the Polchiski-Strominger \cite{Polchinski:1991ax} formalism in \cite{Drummond:2004yp,HariDass:2006sd}  \footnote{ \rev{See also \cite{Drummond:2006su,Dass:2006ud,HariDass:2007dpl,HariDass:2009ub,Dass:2009xe} for a debate on these results and  a discussion on the extension of this analysis to higher orders and its interplay with conformal invariance.}}.} 

In fact, even though the $\SO(D)$ invariance of the original theory is spontaneously broken by the formation of the classical string configuration around which one is expanding, the effective action should still respect this symmetry through a non-linear realization in terms of the transverse fields $X_i$~\cite{Meyer:2006qx, Aharony:2011gb, Gliozzi:2011hj, Gliozzi:2012cx, Meineri:2013ew}. These non-linear constraints induce a set of recursive relations among the coefficients of the expansion, which strongly constrain the coefficients $c_i$. In particular, it can be shown that the terms with only first derivatives coincide with the Nambu-Goto action to all orders in the derivative expansion~\cite{Aharony:2010cx} and that the first correction with respect to the Nambu-Goto action appears at order $1/R^7$ in the large $R$ expansion.
This explains why the Nambu-Goto model has been so succesfull over these last forty years to describe the infrared behaviour of confining gauge theories despite its simplicity and why the deconfinement temperature predicted by Nambu-Goto is so close to the one obtained in Monte Carlo simulations.

This argument can be better understood looking at the original string action, before fixing the reparametrization invariance from a geometric point of view. In this framework the effective action is obtained by the mapping 
\begin{equation}
X^\mu : \mathcal{M} \rightarrow \R^D, \qquad \mu = 0, \cdots , D-1
\end{equation}
of the two-dimensional surface describing the worldsheet of the string $\cal M$ into 
the (flat) $D$-dimensional target space $\R^D$ of the gauge theory  and then imposing the constraints due to Poincar\'e and parity invariance of the original theory. This approach was discussed in detail in ref.~\cite{Aharony:2013ipa}. The first few terms of the action compatible with these constraints must be combinations of the geometric invariants which can be  constructed from the induced metric $g_{\alpha\beta}=\partial_\alpha X^\mu\partial_\beta X_\mu$. These terms can be classified according to their ``weight'', defined as the difference between the number of derivatives minus the number of fields $X^\mu$ (i.e., as their energy dimension). Due to invariance under parity, only terms with an even number of fields should be considered. The first term of this expansion, which is also the only term of weight zero, corresponds, as we mentioned above, to the Nambu-Goto action
 \begin{equation}
 \SNG=\sigma_0\int d^2\xi \sqrt{g}~,
 \end{equation}

At weight two, two new contributions appear:
\begin{eqnarray}
&& S_{2,\cal{R}}=\gamma\int d^2\xi \sqrt{g} \cal{R}, \\
&& S_{2,K}=\alpha\int d^2\xi \sqrt{g} K^2, \label{ext}
\end{eqnarray}
where $\alpha$, and $\gamma$ are two new free parameters, $\cal{R}$ denotes the Ricci scalar constructed from the induced metric, and $K$ is the extrinsic curvature, defined as $K=\Delta (g) X$, with  
\begin{equation}
\Delta(g)=\frac{1}{\sqrt{(g)}}\partial_a[\sqrt{(g)}g^{ab}\partial_b]
\end{equation}
the Laplacian in the space with metric $g_{\alpha\beta}$.
In principle the new free parameters $\alpha$, and $\gamma$ should be fixed, as we did for $\sigma_0$ by comparing with Monte Carlo simulations.
However this process is simplified by the observation that the term proportional to $\cal{R}$ is a topological invariant in two dimensions and, since in the long-string limit in which we are interested one does not expect topology-changing fluctuations, its contribution can be neglected~\cite{Aharony:2013ipa}. On the other hand, the term in eq.~(\ref{ext}) which contains $K^2$ leads to quantum corrections which decrease exponentially with the interquark distance \cite{Caselle:2014eka} and are thus negligible unless the ratio between the coefficient of the $K^2$ term and the string tension grows to infinity in the continuum limit and this seems to occur only in very few models like, for instance, the $d=3$ $U(1)$ model \cite{Caselle:2014eka} . \rev{In these cases an Effective String Theory model, which combines Nambu-Goto and extrinsic curvature was proposed long ago in  \cite{Polyakov:1986cs,Kleinert:1986bk}. The resulting EST is usually known as "rigid string". We shall comment on this issue in the last section of the review.} 

At weight four, two new combinations can be constructed and correspondingly two new parameters\footnote{Notice however that also these new parameters are not compeletly free and can be constrained
 using a bootstrap type of analysis \cite{EliasMiro:2019kyf} in the framework of the S-matrix approach pioneered by \cite{Dubovsky:2012sh}.} appear, leading in the open string channel (i.e. in the low $T$ regime) to the $1/R^7$ correction mentioned above. 
 As above they should in principle be fixed by comparing with Monte Carlo estimates of the potential, however their contributions appear at such a high level that they are very difficult to detect even with the most precise numerical simulations. 
 
\subsection{Beyond Nambu-Goto: the boundary term}
\label{sect:boundary}
Another term which must be considered beyond the Nambu-Goto one is the so called "boundary term".  This term has an origin different from those discussed above.  It is due to the presence of the Polyakov loops at the boundary of the correlator. The classical contribution associated to this correction is the constant term $c$  which appears in the potential and that we have systematically neglected in the previous analysis.
Beyond this classical term we may find quantum corrections due to the interaction with the flux tube. 
The main result in this context is that also these terms are strongly constrained by Lorentz invariance

The first boundary correction compatible with Lorentz invariance is \cite{Billo:2012da}
\eq
b_2\int d\xi_0 \left[
\frac{\partial_0\partial_1 X\cdot\partial_0\partial_1 X}{1+\partial_1 X\cdot\partial_1X}-
\frac{\left(\partial_0\partial_1 X\cdot\partial_1 X\right)^2}
{\left(1+\partial_1 X\cdot\partial_1X\right)^2}\right]\,.
\label{firstb}
\en 
with an arbitrary, non-universal coefficient $b_2$.
The lowest order term of the expansion of eq.(\ref{firstb}) is:

\begin{equation}
\label{derexpsb1}
S_{b,2}^{(1)} = b_2\int d\xi_0 
(\partial_0\partial_1 X)^2~
\end{equation}

The contribution of this term to the interquark potential was evaluated in \cite{Aharony:2010cx} using the zeta function regularization:
\eq
\label{polybound}
\langle S^{(1)}_{b,2} \rangle=-b_2\frac{\pi^3 N_t}{60 R^4} E_4(e^{-\frac{\pi N_t}{ R}})
\en
where $E_4$ denotes the fourth order Eisenstein series (see the Appendix for definitions and properties of these functions). In the standard low temperature ($N_t\gg R$) 
setting, this amounts to a correction proportional to $1/R^4$ to the interquark potential, which turns out to be the dominant correction term beyond Nambu-Goto in this regime and represents a further  obstacle to detect signatures of the ``bulk'' correction terms discussed in the previous subsection.

Recent high precision Monte Carlo simulations \cite{Brandt:2010bw,Brandt:2017yzw,Brandt:2018fft,Brandt:2021kvt,Billo:2012da} allowed to estimate $b_2$  for a few LGTs with remarkable precision. For the $SU(2)$ model in (2+1) dimensions a boundary correction was estimated even for the first string excitations \cite{Brandt:2021kvt}. Preliminary results have been also obtained for the (3+1) dimensional SU(3) LGT \cite{Bakry:2020ebo,Bakry:2019cuw} where, besides $b_2$, a tentative estimate of the next to leading term $b_4$  is also reported.
As expected these values are not any more universal and represent the first hint of the fact that different LGTs are described by different ESTs and that the information on the gauge group of the model and the gluon content of the flux tube is somehow encoded in the effective string model.

At the same time the above discussion shows that this boundary term is the dominant non universal correction 
beyond Nambu-Goto in this low $T$ regime and it is clear that its presence makes it almost impossible
to detect the much weaker signatures of the ``bulk'' correction terms discussed in the previous subsection.

However, by performing a modular transformation (see the Appendix \ref{AppendixA}) it is easy to see that in the high temperature limit (i.e. $R\gg N_t$) 
this correction becomes 

\eq
\label{polybound2}
\langle S^{(1)}_{b,2} \rangle=-b_2\frac{4 \pi^3}{15 N_t^3} E_4(e^{-\frac{4\pi R}{N_t}})
\en
and does not contain a term proportional to $R$ and thus it does not give a correction to the temperature dependent string tension $\sigma(T)$!

This is a second important reason of interest of the high temperature regime which is the focus of this review.  In this limit the boundary term does not interfere with the "bulk" EST corrections beyond Nambu-Goto which can thus be directly observed with
 Monte Carlo simulations.

\section{Comparison with Monte Carlo Simulations}

In the past years the predictions of EST for the interquark potential were tested with Monte Carlo simulations of increasing precision in several different LGTs
\cite{{Hasenbusch:1992zz},{Caselle:1996ii},Ambjorn:1984me,Ambjorn:1984yu,Necco:2001xg,
{Juge:2002br},{Lucini:2001nv},{Luscher:2002qv},{Caselle:1994df},{Lucini:2002wg},{Caselle:2004er},{Caselle:2005xy},{Bringoltz:2006zg},{Caselle:2006dv},{HariDass:2006pq},{Bringoltz:2008nd},{Athenodorou:2011rx},
{Athenodorou:2010cs},{Caselle:2010pf},{Billo:2011fd},{Mykkanen:2012dv},{Athenodorou:2013ioa},{Caselle:2016wsw},{Athenodorou:2016kpd}}. 
Most of these tests were performed in the low temperature regime. 
However, as we have seen in the previous sections, in order to have a complete understanding of EST and to test its consistency under the open-closed string transformation, it is interesting to test EST predictions also in the high temperature regime. This is the goal of this section
in which we shall report the results of a few papers in which EST was compared with high-$T$ Monte Carlo simulations.

We shall first discuss in detail, as an example, the $SU(2)$ gauge theory in $(2+1)$ dimensions, which is the simplest non-abelian LGT  and allows to reach high precision results with a relatively small amount of computing power. Then, in the last subsection, we shall briefly review the results obtained in other LGTs both in (2+1) and in (3+1) dimensions.

\subsection{LGT observables}
\label{observables_subsection}

Let us first discuss a few combinations of Polyakov loop correlators which are particularly useful to address the comparison between EST predictions and LGT results in the high temperature regime.

To simplfy notations let us define the Polyakov loop correlator as:
\eq
G(R,T)=\langle P(x)P^\dagger(x+R) \rangle_{N_t} 
\label{polya2}
\en

Following ~\cite{Luscher:2002qv},~\cite{Caselle:2004er} and \cite{Caselle:2011vk}
it is particularly convenient to introduce the following quantities:
\eqa
\label{Q_definition}
Q(R,T) &=& {T} \ln \frac{G(R,T)}{G(R+1,T)}, \\
\label{A_definition}
A(R,T) &=& {R^2} \ln \frac{G(R+1,T)G(R-1,T)}{G^2(R,T)}.
\ena
Note that, in the continuum limit $a \to 0$, $Q(R,T)$ tends to the first derivative of $V(R,T)$ with respect to $R$:
\eq
\label{continuum_limit_of_Q}
\lim_{a \to 0} Q = \frac{\partial V}{\partial R},
\en
so that it can be interpreted as a lattice version of (minus) the interquark force. On the other hand, $A(R,T)$ is a dimensionless quantity proportional to the discretized derivative of the force:
\eq
\label{continuum_limit_of_A}
\lim_{a \to 0} A = - \frac{R^2}{T} \frac{\partial^2 V}{\partial R^2}.
\en

These quantities are the finite temperature version of the observables introduced in~\cite{Luscher:2002qv}. In particular
 $Q(R,T)$ coincides in the low-$T$ limit with the ``force'' $F(R)$ of~\cite{Luscher:2002qv} 
while $A(R,T)$ is related to the ``central charge'' $c(R)$  of~\cite{Luscher:2002qv}  as follows
\eq
A(R,T)=\frac{2}{RT}c(R).
\en

Using eq.~(\ref{potential_D=3}) we may estimate the large-$R$ limit of 
these two observables for $D=3$ LGTs in the framework of the
Nambu-Goto effective string model:
\eqa
\label{NG_prediction_for_Q} Q(R,T) &\simeq& \sigma(T) + \frac{T}{2R} - \frac{T^2}{8 \sigma(T) R^2} + \cdots \\
\label{NG_prediction_for_A} A(R,T) &\simeq& \frac{1}{2} - \frac{T}{4 \sigma(T) R} +\cdots
\ena

The constraints on the EST discussed in sect.(\ref{beyond}) tell us that these expressions for $Q$ and $A$ should be universal and should hold for any LGT (except, as usual, the 3d $U(1)$ LGT).  Corrections to the EST beyond Nambu-Goto should only affect higher order terms 
(the dots in the above eq.s (\ref{NG_prediction_for_Q},\ref{NG_prediction_for_A}),  and are expected to affect the finite temperature string tension $\sigma(T)$ only with corrections of the order of $(T/T_c)^7$. In the next section we shall compare this prediction with Monte Carlo simulations.

\subsection{The $SU(2)$ LGT in $(2+1)$ dimensions.}

The (2+1) dimensional $SU(2)$ model has been the subject of several numerical efforts in the last years, 
 most of them however focused on the low $T$ regime of the model. We shall report here the results of the simulations 
discussed in \cite{Caselle:2011vk} which were instead performed at a relatively high ($T=\frac34 T_c$) temperature.

The only imput we need to fix our predictions is the zero temperature string tension $\sigma_0$.  This can be fixed using for instance the results of ~\cite{Caselle:2004er} which we report here
\eq
\label{sigma_a2}
\sqrt{\sigma_0 }\simeq  
 \frac{1.324(12)}{\beta} + \frac{1.20(11)}{\beta^2}  
\en 

Simulations were perfomed at $\beta=9$ for which we have $\sigma_0=0.0262(1)$ \cite{Caselle:2004er} on a lattice of size $120^2 \times 8$. For this value of $\beta$ the critical tempearture is, almost exactly located at $N_t=6$ thus this choice of lattice sizes corresponds to a temperature $T=\frac34 T_c$. Polyakov loop correlators were measured up to the distance of $R=19$ lattice spacings.
We report for completeness the results of the simulations in Table~\ref{SU2_results_table} and refer the interested reader to 
~\cite{Caselle:2004er} for more details on the simulation settings and on the fitting protocol.

\begin{table}[ht]
\centering
\phantom{-------}
\begin{tabular}{|cc|cc|cc|}
\hline
 $R$ & $ Q$ &  $R$ & $Q$ &  $R$ & $Q$ \\
\hline
   2 & 0.037433(46)  &      8 & 0.02232(11)   &        14 & 0.01971(16)   \\
   3 & 0.030958(56)  &      9 & 0.02170(12)   &        15 & 0.01949(18)    \\
   4 & 0.027600(64)  &     10 & 0.02117(12)   &        16 & 0.01926(19)    \\
   5 & 0.025553(72)  &     11 & 0.02072(13)   &        17 & 0.01906(20)    \\
   6 & 0.024154(84)  &     12 & 0.02034(15)   &        18 & 0.01892(22)    \\
   7 & 0.023118(94)  &     13 & 0.02000(15)   &        19 & 0.01876(24)    \\
\hline
\end{tabular}
\phantom{-------}
\caption{Results for $Q(r,T)$, as a function of the interquark distance $R$, for the (2+1) dimensional $\SU(2)$ model at $T=3T_c/4$, taken from \cite{Caselle:2011vk}. }
\label{SU2_results_table}
\end{table}

   Following eq.(\ref{NG_prediction_for_Q}) the values of $ Q(R,T) $ are fitted with:
\eq
\label{SU2_Q_fit}
 Q (R,T)|_{T=3T_c/4} = s + \frac{b}{R} + \frac{c}{R^2} \, , 
\en
   and the following best fit values for the parameters are found
   \begin{equation} \nonumber
    s= 0.01530(37)\ 
    b= 0.0668(58)\
    c= - 0.087(27)
   \end{equation}
   with a reduced $\chi^2_r=0.75$.

   The universal correction in which we are interested are encoded in the parameter $b$ which according to the analysis discussed in the previous sections should be given
   by
   \begin{equation} \nonumber
    b=\frac{T}{2}=\frac{1}{16}=0.0625 \, ,
   \end{equation}
   which turns out to be in remarkable agreement with the result of the fit.
   
   This is further confirmed by the analysis of the $A(R,T)$ values (which can be easily obtained from the data reported
   in table~\ref{SU2_results_table}). These values are fitted with
\eq
\label{SU2_A_fit}
A (R,T)|_{T=3T_c/4} = k - \frac{m}{R} \, , 
\en
   finding
   \begin{equation} \nonumber
    k=0.528(28),\;\; 
    m=-1.09(28), \;\;\; \mbox{with  $\redchisq=1.6$},
   \end{equation}
   which is again in perfect agreement with the expected value $k=1/2$. 
   
   From the first fit we can extract the value $\sigma(T)=0.01530(37)$ for the finite temperature string tension at $T=1/8$.  
   Using the value $\sigma_0$ reported in eq.~(\ref{sigma_a2}), we may obtain a ``Nambu-Goto'' prediction for $\sigma(T)$ using 
   eq.~(\ref{sigma_T}), which turns out to be
   $\sigma_{\mbox{\tiny{NG}}}\left(T=1/8\right)=0.01605(6)$, at two standard deviations from the observed value. 
   This indicates, as already observed in \cite{Athenodorou:2007du},
   that for the (2+1) $\SU(2)$ LGT the Nambu-Goto string represents a rather good approximation but, as the precision of the simulations improves, small deviations start to be detected. These deviations are the signatures of the $(T/T_c)^7$ term mentioned above.
   
   Finally, using the measured value of $\sigma(T)$ it is possible to obtain predictions for the subleading corrections in the two fits.
   One find for the $c$ term in the first fit    $c_{\mbox{\tiny{NG}}} \sim -0.1216(5)$ and for $m$ in the second fit $m_{\mbox{\tiny{NG}}}= -1.946(8)$. Both values are similar to those extracted from the fits, but  not
   compatible within the errors. 
   This small discrepancy agrees in sign and magnitude with the analogous deviations from the Nambu-Goto ansatz observed in \cite{Athenodorou:2007du} and summarized in the 
   coefficient $C_3$ evaluated there. We shall comment on these deviations in the next section.

\subsection{EST predictions versus Monte Carlo results for different LGTs.}

The same analysis was performed in \cite{Caselle:2004er} for the $(2+1)$ $SU(3)$ and $SU(4)$ lattice gauge theories and, using data obtained in \cite{Caselle:2002ah}, also in the case of the three dimensional 
Ising gauge model for two different temperatures. We summarize the results for the fits to $Q(R.T)$ in Table~\ref{all_results}.

\begin{table}[h]
\hskip-1cm
\phantom{-------}
\begin{tabular}{|c|c|c|c|c|c|c|c|c|}  
\hline
 gauge group & $N_t$ &  $T/T_c$& $s$ & $\sigma_{\mbox{\tiny{NG}}}(T)$ & $b$ & $b_{\mbox{\tiny{NG}}}$ &  $c$ & $c_{\mbox{\tiny{NG}}}$  \\
\hline
$\SU(2)$ & $8$ &  $3/4$ & $0.01530(37)$  & $0.01605(6)$ & 0.0668(58)& 0.0625 &  $-0.087(27)$ & $-0.1216(5)$ \\
$\SU(3)$ & $8$&  $3/4$  & $0.01884(44)$  & $0.01946(6)$ & 0.0612(74)& 0.0625 &  $-0.063(34)$ & $-0.1003(5)$\\
$\SU(4)$ & $8$&  $3/4$  & $0.01721(43)$  & $0.01830(40)$ & 0.0634(70)& 0.0625 &  $-0.063(32)$ & $-0.1070(20)$\\
$\Z_2$ & $12$&  $1/2$  & $0.01485(2)$  & $0.01487(6)$ & 0.0414(8)& 0.04167 &  $-0.049(6)$ & $-0.058(1)$\\
$\Z_2$ & $9$ &  $2/3$  & $0.01137(11)$  & $0.01067(6)$ & 0.0522(40)& 0.0556 &  $-0.076(34)$ & $-0.145(1)$\\

 \hline
\end{tabular}
\phantom{-------}
\caption{ Results of the fits to $Q(R,T)$ for various LGTs (listed in the first column), together with the expected values for the best fit parameters according to the Nambu-Goto EST.}
\label{all_results}
\end{table}

Looking at the table we can make a few interesting observations
\begin{itemize}
\item
All the models,  except the one at the lowest temperature, show deviations in the fitted value of $\sigma(T)$ with respect to the Nambu-Goto prediction.  These deviations are the signatures of the terms beyond Nambu-Goto which must be included in the EST action which we discussed in the previous section. They are exactly those needed to match the critical index of the deconfinement transition which in this case is $\nu=1$
instead of the Nambu-Goto value $\nu=1/2$. 
\item
The universal constant $b$ is always compatible with the theoretical expectation. This represents a remarkable consistency check of the whole EST construction.
\item
The constant $c$ shows the same trend for all the models:  it is similar to the expected Nambu-Goto value, 
but always slightly smaller in magnitude.
Most likely this deviation is due to the fact that in the fit we are neglecting higher terms, and indeed the first of them, the one
proportional to $1/R^3$ , due to the expansion of the modified Bessel function has the opposite sign with respect to the $1/R^2$ one and may explain the decrease in magnitude of $c$
\end{itemize}

Similar results are found fitting the $A(R,T)$ function (see \cite{Caselle:2004er} for further details).

A similar analysis was also performed for the SU(2) model in (3+1) dimensions \cite{Bonati:2011nt} and (with a different set of observables) 
 for SU(3) in \cite{Bakry:2020flt}. In both cases two different temperatures were tested and  a good agreement with the Nambu-Goto  predictions was found for the lower one, while deviations were detected for the higher one,  pointing to the possible presence of terms in the EST beyond the Nambu-Goto one. Besides its physical relevance, this extension to (3+1) dimensions
 is also interesting because the interquark potential, as can be seen in eq.(\ref{potential_approximate_form}), shows a non-trivial dependece on the number of space time dimensions which is precisely confirmed by the numerical simulations at the lowest temperatures. 
 
As a matter of fact this type of corrections in the (3+1) dimensional SU(3) models were already observed more than twenty years ago when the first high precision determinations of $\sigma(T)$ were obtained~\cite{Kaczmarek:1999mm,Cardoso:2011hh}. The behaviour of $\sigma(T)$ was very similar to the one predicted by the Nambu-Goto action, but  with small deviations in the vicinity of the deconfinement point. Thes deviations led to a non-zero, even if small,
 value of the string tension $\sigma(T_c)$ at the critical point which had the effect of transforming the second order phase transition predicted by the Nambu-Goto model into the first order deconfinement  phase transition of the SU(3) (3+1) dimensional model.

\section{Width of the confining flux tube at high temperature}
\label{sect:width}
One of the most intriguing features of the EST picture of confinement 
is the logarithmic increase of the square width $w^2(R)$ of the
 flux tube as a function of the interquark distance $R$~\cite{Luscher:1980iy}.  
\eq
\begin{array}{c}
\displaystyle
\sigma_0 w^2(R)=\frac{1}{2\pi} \log\frac{ R}{R_c}
\end{array}
\label{ris1}
\en
where $R_c$ is known as "intrinsic width" and sets the scale of the logarithmic growth.

This logarithmic growth, which is commonly referred to as the  "delocalization" of the flux 
tube  was discussed for the first time many years ago by 
 L\"uscher, M\"unster and Weisz in~\cite{Luscher:1980iy} but it required several years of efforts before it could be observed 
 in lattice simulations. The first numerical results were obtained in abelian models \cite{Caselle:1995fh, Zach:1997yz, Koma:2003gi, Panero:2005iu, Giudice:2006hw,Amado:2013rja,Amado:2012wt} where, thanks to duality, simulations can be performed more easily
 and later the flux tube width was studied also in non abelian LGTs~\cite{Bakry:2020ebo,Bakry:2019cuw,Gliozzi:2010zv,Bakry:2010zt,Cardoso:2013lla,Bicudo:2017uyy,Cardaci:2010tb,Cea:2012qw,Cea:2014uja,Cea:2015wjd,Cea:2017ocq,Baker:2018mhw,Baker:2019gsi}. 
 
 An important issue in this context is to understand the fate of the flux tube  width as 
 the deconfinement transition is approached from below. It is important to stress that delocalization and deconfinement are two deeply different conditions of the flux
 tube. As we have seen in sect.\ref{sect:LGT}, deconfinement is characterized by the vanishing of the string tension $\sigma(T)$ and, accordingly, of the flux tube. 
 The delocalization of the flux tube instead coincides with the onset of the rough phase.   Delocalization is a typical quantum
 effect. It is a consequence of the Mermin-Wagner theorem which imposes the restoring in the continuum limit 
 of the translational symmetry for the fluctuations of the flux tube in the transverse directions. Intutively it amounts to say that we cannot fix deterministically 
 the trajectory of the flux tube but may only describe it as a probabilty distribution.
 It is important to stress that, even if
 delocalized, the flux tube fully keeps its confining function. The quantum fluctuations which drive the delocalization also influence the confining potential
 (as the presence of L\"uscher term indicates) but do not destroy it. 
  
 While the behaviour of the string tension $\sigma(T)$ as the deconfinement temperature $T_c$ is approached from below is rather well understood, much less is  known on the behaviour of the flux tube thickness in this regime.
 This is an important issue from a physical point of view 
 since the interplay between delocalization and deconfinement could strongly influence the transition from hadrons to free quarks as $T_c$ is approached.

 Similarly to what we did for the interquark potential,  also this problem can be addressed by performing a modular 
 transformation of the low temperature result. This was done in~\cite{Allais:2008bk} in the case of the free Gaussian action 
 (i.e. the first order in the perturbative expansion of the Nambu-Goto effective string) leading in the large $R$ limit to the following result:

\eq
  \sigma_0 w^2_{lo}=\frac{R}{4N_t} + \frac{1}{2\pi} \log\frac{N_t}{L_c} -\frac{1}{\pi}e^{-2\pi\frac{R}{N_t}}+\cdots
\label{risfinale}
\en 

where $L_c$ is a length scale which plays the role in this limit of the intrinsic width of eq.(\ref{ris1}) and the suffix $lo$ is added to emphasize that this is only the leading order (Gaussian) approximation of the true flux tube width.
 
  We see that the large $R$ behaviour of the square width changes
 completely and becomes linear (with a 
coefficient $\frac{1}{4\sigma_0 N_t}$) instead of logarithmic. 
This behaviour holds in principle for any temperature $T$, but as $T$ decreases it requires larger and larger values
 of $R$ to be observed. Similarly it is possible to show that for any fixed value of $R$  the square width smoothly converges toward the expected 
 logarithmic behaviour as $T$ decreases. The threshold between the two
 behaviours is  $R\sim 1/T$. 
 
 In~\cite{Allais:2008bk} this prediction was tested with a set of high precision  Monte Carlo simulations of the 3d gauge Ising models 
 and  only a partial agreement with eq.(\ref{risfinale})  was found. 
 For all the temperatures studied in~\cite{Allais:2008bk} $w^2(R)$ was indeed 
 a linearly increasing function of $R$. However the coefficient of this linear behaviour was in general larger
 than the one predicted by the effective string (except for the smallest temperature values) 
 and, what is more important, it seemed to diverge as the deconfinement point was approached (while the coefficient $\frac{1}{4\sigma_0 N_t}$ converges instead to a finite value at the deconfinement point).
 
 This discrepancy tells us that as the deconfinement transition is approached the leading order approximation gets worse and worse and that, similarly to what happens for the interquark potential, higher order terms must be included.
 The problem is that, while for the interquark potential we have the exact solution to all orders, for the flux tube width only the next to leading order is known ~\cite{Gliozzi:2010zv,Gliozzi:2010zt,Gliozzi:2010jh}.  This correction goes in the right direction but is not enough to fill the gap between numerical data and theoretical expectations.

We shall see in the next section that the Svetitsky-Yaffe conjecture offers a powerful tool to address this issue when one approaches the deconfinement transition and allows to guess the resummation to all orders of the flux tube width for the Nambu-Goto effective string.
The complete answer for the leading term linear in $R$ turns out to be 
\cite{Caselle:2006wr,Caselle:2010zs,Caselle:2012rp}

\eq 
w^2(R)\, = \,  \frac{1}{4\sigma(T)} \, {R}{T} 
\label{sy3bis}
\en 
where $\sigma(T)$ is the temperature dependent string tension of eq.(\ref{sigma_T}).

By expanding this expression in powers of $T/T_c$
it is easy to see that both the leading order $w^2_{lo}$ and the next to leading order 
of ~\cite{Gliozzi:2010zv,Gliozzi:2010zt,Gliozzi:2010jh}
fully  agree with eq.(\ref{sy3bis}).

The results for the Ising model of~\cite{Allais:2008bk} agree with eq.(\ref{sy3bis}) and a few years later, the same behaviour was observed with a set of high precision simulations in the $(3+1)$ dimensional $SU(3)$ model \cite{Bicudo:2017uyy}.

To understand the origin of eq.(\ref{sy3bis}) we should first define the LGT observables which allows to evaluate the flux tube width, then address their dimensionally reduced version,  according to the Svetitsky-Yaffe projection and finally evaluate these expectation values using the S-matrix approach.  Let us address these issues step by step.
 
\subsection{Definition of the flux tube thickness} 

In a finite temperature setting the lattice operator which is used to evaluate the flux through a plaquette $p$ of the lattice is:
\eq 
\bra\phi(p;P,P')\ket_{N_t}=\frac{\bra P P'^\dagger~U_p\ket_{N_t}}{\bra PP'^\dagger \ket_{N_t}}-\bra U_p\ket_{N_t }
\label{flux2} 
\en 
where $P$, $P'$ are two Polyakov loops separated by $R$ lattice spacings and $U_p$ is 
the operator associated with the plaquette $p$.
Different possible orientations of the
plaquette $p$ measure different components of the flux. In the following we shall neglect this dependence which
plays no role in our analysis. The only information that we need is the position of the plaquette.
Let us define 
\[
 \bra\phi(p;P,P')\ket_{N_t }=\bra\phi(\vec h;R,N_t)\ket
\]
where $\vec h$ denotes the displacement of $p$ from the $P$ $P'$ plane. 
In each transverse direction, the flux density 
shows a Gaussian like shape (see for instance Fig. 2 of ~\cite{Caselle:1995fh}). 
The width of this Gaussian $w$ is the quantity which is usually denoted as ``flux tube thickness'':

\eq
w^2(R,N_t)=\frac{\sum_{\vec h} \vec h^2 \bra\phi(\vec h;R,N_t)\ket}{\sum_{\vec h} \bra\phi(\vec h;R,N_t)\ket}
\label{w1}
\en

This quantity depends on the number of transverse dimensions and on the bare gauge coupling $\beta$. 
Once $\beta$ is fixed the only remaining dependences are 
on the interquark distance $R$ and  on the inverse temperature $N_t$. 
By tuning $N_t$ we can thus study the flux tube thickness near the deconfinement transition.

\subsection{Dimensional reduction and the Svetitsky--Yaffe approach.} 
As we have seen in sect.\ref{sysec}
In the vicinity of the deconfinement transition the physics of a $(d+1)$ LGT can be described using an effective model in which the spacelike links
are integrated out and the only remaining degrees of freedom are the Polyakov loops. 
The simplest examples of this effective mapping are 
the (2+1) $SU(2)$ LGT  and the (2+1) Ising gauge model which have the same center $Z_2$ and are thus both mapped into the $2d$ spin Ising model.
We shall use this case as an example in the following to simplify the discussion.
Using the correspondences discussed in sect.(\ref{sysec})
it is possible  to construct the dimensionally reduced projection of the operator which measures the flux tube thickness
which turns out to be a suitable ratio of three and two  point correlators of the spin and energy operators (see~\cite{Caselle:2006wr} for a detailed discussion of this mapping). 
In the particular case of the $2d$ Ising model that we are using as an example this combination is:

\eq 
\frac{\langle \sigma(x_1) \epsilon(x_2) \sigma(x_3)\rangle} 
{\langle \sigma(x_1) \sigma(x_3)\rangle} 
\en 
to be evaluated
 in the high temperature phase and in zero magnetic field. 
Since we are interested in the large distance behaviour of these correlators we can use the so-called Form Factors approach (see 
\cite{Yurov:1990kv} for an introduction to Form Factors and their application in the context of the 2d Ising model without magnetic field). 

A straightforward calculation leads to the following  expression for
the flux  distribution~\cite{Caselle:2006wr}
\eq 
P(R,y) \ = \frac{ 2\pi  R}{4y^2+R^2} \, \frac{e^{-m \sqrt{4y^2+R^2}}}{K_0(mR)}.  
\label{nongaussian} 
\en 
 where $y$ denotes the transverse direction, $K_0$ is the modified Bessel function of order 0,  $m$ is the mass of the 2d Ising model
and a large $mR$ limit is assumed.

From this flux distribution it is easy to extract the square of the flux tube width as the ratio

\eq 
w^2(R) \ = \frac{\int_{-\infty}^{\infty} dy \, y^2 \, P(R,y)}{\int_{-\infty}^{\infty} dy \, P(R,y)}  
\label{sy1}
\en 
which, setting $x=2y/R$ amounts to evaluate
\eq 
w^2(R) \ = \frac{R^2}{4} \, \frac{\int_{-\infty}^{\infty} dx \, \frac{x^2}{1+x^2} e^{-2mr \sqrt{1+x^2}}}{\int_{-\infty}^{\infty} dx \, \frac{e^{-2mr \sqrt{1+x^2}}}{1+x^2} }  
\en 
These integrals can be evaluated asymptotically in the large $mR$ limit \cite{Caselle:2006wr,Caselle:2010zs}
leading to the following result:
\eq 
w^2(R) \simeq \  \frac{1}{4} \, \frac{R}{m} + \dots.
\label{sy2}
\en 
where the dots stay for terms constant or proportional to negative powers of $R$.

The last step in order to compare this result with eq.(\ref{sy3bis}) is to give a meaning to the Ising mass $m$ in terms of LGT quantities.

This can be easily accomplished if we recall that the mass can be obtained from the large $R$ limit of the
spin spin correlator, which according to the Svetitsky-Yaffe mapping is the 2d 
limit of the expectation value of two Polyakov loops at distance $R$.
Following eq.(\ref{E0}) we can thus identify
\eq 
  m=\sigma(T) N_t
\en 

from which we immediately obtain the result of eq.(\ref{sy3bis})

Similar arguments allow to obtain also estimates of the intrinsic width of the model \cite{Caselle:2012rp}.

The above analysis was performed in the case of the Ising model, but the argument is completely general and the derivation of the large distance behaviour holds for any spin model with a gap in the spectrum.

\section{Open issues and concluding remarks}
\label{sec:conclusions}

In this review we focused in particular on the behaviour of the interquark potential and of the flux tube width.  There are however a few other observables which show a non trivial behaviour at high-$T$ and allow for non-trivial tests of EST. We could not discuss them in detail in this review for lack of space and specific expertise but we briefly mention them here and list a few relevant references which may help the interested reader to deepen the subject.

\begin{itemize}

\item
{\bf The deconfinement transition as a Hagedorn transition.}

One of the more interesting consequences of the EST description of confinement is that the deconfinement transition can be interpreted as a Hagedorn transition~\cite{Hagedorn:1965st}. This can be understood (using a dual transformation) as a direct consequence of the tachyonic singularity in the interquark potential~\cite{Olesen:1985ej}. This Hagedorn behaviour has relevant consequences on the equation of state of pure gauge theories 
which can be precisely tested using Monte Carlo simulations. In fact, in pure gauge theories the only massive excitations in the confining phase are glueballs and the equation of state can be accurately modeled in terms of a gas of these massive, non-interacting glueballs.
If one assumes a description of glueballs  as  closed color flux tubes (as for instance in the Isgur-Paton model~\cite{Isgur:1984bm}) then one should expect a Hagedorn-like \cite{Hagedorn:1965st} stringy behaviour of the glueball spectrum and as a consequence a highly non trivial temperature dependence of pressure and entropy across the deconfinement transition.  This effect was observed for the first time in ref.~\cite{Meyer:2009tq} for the $\SU(3)$ Yang-Mills theory in (3+1) dimensions, and later also in  $\SU(N)$ theories in (2+1) dimensions~\cite{Caselle:2011fy} and in the (2+1) dimensional $SU(2)$ \cite{Caselle:2015tza,Alba:2016fku}, finding  always a very good agreement with the expected Hagedorn behaviour.

\item {\bf The spacelike string tension at high Temperature.}

An interesting open issue in Lattice Gauge Theory is to understand and model the behaviour of the so called ``space--like string tension" \cite{Karkkainen:1993ch,Bali:1993tz,Karsch:1994af,Caselle1994f,Koch:1994zt,Ejiri:1995gd,Sekiguchi:2016gxx,Schroder:2005zd,Liddle:2007uy,Cheng:2008bs,Maezawa:2007fc,Alanen:2009ej,Andreev:2006eh,Andreev:2007rx,Meyer:2005px} across the deconfinement transition. 

The space--like string tension is extracted from the correlator of space--like Polyakov loops, i.e. Polyakov loops which lay in a space--like plane, orthogonal to the compact time direction $N_t$. Due to their space--like nature these Polyakov loops  do not play the role of order parameter of deconfinement and the space--like  string tension extracted from them is different from the actual string tension of the model $\sigma(T)$.

At low temperature the two string tensions coincide but as the temperature increases they behave differently \cite{Karkkainen:1993ch,Bali:1993tz,Karsch:1994af,Caselle1994f,Caselle:1993cb}. As we have seen $\sigma(T)$ decreases as the deconfinement temperature is approached and vanishes at the deconfinement point, while the space--like string tension remains constant and then increases in the deconfined phase \cite{Karkkainen:1993ch,Bali:1993tz,Karsch:1994af}.  The physical reason for this behavior is that the correlator of two space--like Polyakov loops describes quarks moving in a finite temperature environment. It can be shown that what we called space--like string tension is related to the screening masses in hot QCD \cite{Koch:1994zt,Ejiri:1995gd,Sekiguchi:2016gxx,Schroder:2005zd,Liddle:2007uy,Cheng:2008bs,Maezawa:2007fc} and thus it does not vanish in the deconfined phase.  

An EST description of this behaviour has been recently obtained \cite{Beratto:2019bap} using the mapping between the Nambu-Goto action and the
 $T\bar T$ deformation of the free bosonic action. An important open issue in this context is to address the interplay of the space--like string tension with the intrinsic width of the flux tube.

\item
{\bf EST and interfaces.}

In this review we studied EST in two particular choices of boundary conditions for the world sheet:  Wilson loops (rectangular geometry) and Polyakov loop correlators (cylindrical geometry).  There is a third important case, the toroidal geometry, which cannot be easily realized in non-abelian LGTs, but is pretty natural in three dimensional abelian gauge theories.  
These models, thanks to the Kramers--Wannier duality can be mapped into standard three dimensional spin models (the most relevant example being the 3d gauge Ising model which is mapped into the three dimensional Ising spin model). By suitably choosing the boundary conditions of the spin model (for instance: antiperiodic in the Ising case) in the low temperature phase one can induce the formation of interfaces 
which can be described by EST with a toroidal world sheet \cite{Munster:1990yg,Caselle:1992ue,Klessinger:1992qq,Caselle:1994df,Hoppe:1997ey, Muller:2004vv,Caselle:2006dv,Caselle:2007yc}. Interfaces in the spin model are in some sense the dual of the Wilson loops in the gauge model.
The partition function of the Nambu-Goto string with this toroidal boundary conditions can be evaluated with the same tools used for the Polyakov loop correlators \cite{Billo:2006zg}. The major reason of interest of this setting is the absence of boundary terms. It is thus much easier to study higher order terms of EST and in fact some of the most precise Monte Carlo studies of these terms were obtained using interfaces in the 3d Ising model \cite{Caselle:2007yc,Caselle:2016wsw}.
The analogy of the high temperature regime in this context is obtained by  "squeezing" the interface in one direction. From the spin model point of view this is the regime in which one is approaching dimensional reduction from three to  two dimensions \cite{Billo:2007fm}.  A systematic comparison of EST predictions and Monte Carlo simulations in this regime is still lacking and could lead to an interesting and original insight into EST behaviour.

\item 
{\bf Interplay between the EST and the dual superconductor model of confinement}

In this review we introduced the EST, following the seminal papers of L\"uscher and collaborators, as a tool to describe the behaviour of Wilson loops in LGT beyond the roughening transition.  There is however a different, interesting, route which may lead to an effective string description of confinement which was proposed long time ago by Nielsen and Olesen~\cite{Nielsen:1973cs}, 't Hooft~\cite{'tHooft:1974kcl}, Mandelstam~\cite{Mandelstam:1974pi} and Polyakov~\cite{Polyakov:1974ek}. The proposal relies on the description of the QCD vacuum as a coherent state of color magnetic
monopoles or, equivalently, as a  
magnetic (dual) superconductor (for a review see for instance \cite{Ripka:2003vv,Antonov:2004kj,Antonov:2017lur}). 
According to this picture the (dual) Meissner effect naturally leads to vortex like structures: the Abrikosov vortices~\cite{Abrikosov:1956sx} which are very similar to the confining color flux tubes which are described by the EST. 
A very interesting laboratory to address this picture is the 3d $U(1)$ LGT for which it can be shown, using a duality transformation, that confinement is indeed due to the condensation of monopoles~\cite{Polyakov:1976fu}. The remarkable success of this approach led to conjecture that a similar mechanism could drive confinement also in non-Abelian Yang-Mills theories~\cite{Cardaci:2010tb,Cea:2012qw,Cea:2014uja,Cea:2015wjd,Cea:2017ocq,Baker:2018mhw,Baker:2019gsi}.

The implicit assumption behind this scenario is that there should exist a duality transformation mapping gauge fields into strings. In the non-Abelian case, such gauge/string duality transformation is in general unknown,\footnote{A notable exception, however, is given by the holographic correspondence, relating gauge theories and string theories defined in a higher-dimensional spacetime~\cite{Maldacena:1997re, Gubser:1998bc, Witten:1998qj}.} but in the 3D $U(1)$ case Polyakov~\cite{Polyakov:1996nc} (see also~\cite{Antonov:1998kw,Antonov:2004kj,Antonov:2017lur} for an alternative derivation) was able to give a heuristic proof of this mapping and proposed to describe the free energy of a large Wilson loop with a string action combining both the Nambu-Goto and the extrinsic curvature terms, the so called ``rigid string'' \cite{Polyakov:1986cs,Kleinert:1986bk}. 

It is by now clear that this approach leads to an EST {\sl different} from the one discussed in this review~\cite{Caselle:2014eka}. The "rigid string", dominated by the extrinsic curvature term, agrees with the expectation of the dual superconductor model while the one which we discussed in this review has a negligible  extrisic curvature term and is dominated by the Nambu-Goto behaviour. The major difference between the two ESTs is in the shape and width of the flux tube~\cite{Caselle:2016mqu}.  Interestingly this difference is magnified exactly in the high temperature regime~\cite{Caselle:2016mqu,Caselle:2019khe} which is the subject of this review. It would be interesting to pursue this study to better understand the role of the extrinsic curvature term in driving this difference and, more importantly, which one better describes the behaviour of the flux tube in non-abelian LGTs.

\rev{As a final remark on this issue, let us stress that the rigid string shows a pretty different behaviour depending on the sign of the extrinsic curvature term. An EST with negative extrinsic curvature was proposed more than twenty years ago in  \cite{Orland:1994qt,Sato:1994vz,Kleinert:1996au} and was subsequently thoroughly  studied in \cite{Antonov:2004kj,Antonov:2017lur,
Diamantini:1998zu,Diamantini:2002rp,Diamantini:2002wt,Hidaka:2009xh}. Despite the apparent instability due to the negative sign of the curvature term, it can be shown that the string is stabilized by higher order terms in the derivative expansion \cite{Diamantini:1998zu} (for a review, see for instance  \cite{Antonov:2004kj}). In particular, as far as the topic of this review is concerned, the high temperature behaviour of the model was studied in detail in \cite{Diamantini:2002rp,Diamantini:2002wt} and, also in this case, it would be very interesting  to test these prediction with high precision Monte Carlo data for non-abelian LGTs.}

\end{itemize}

In the last few years we have witnessed a remarkable progress in our understanding of EST, however several important issues are still open,  from the identification of EST terms beyond the Nambu-Goto one,  to a better understanding of the role and properties of the rigidity term. The main goal of this review was to show that the high-$T$ regime of LGTs is a perfect laboratory to test new ideas in this context and 
compare them with Monte Carlo simulations. We hope that this review will stimulate further research in this direction.

\vskip 1.5cm
\noindent {\large {\bf Acknowledgments}}
\vskip 0.2cm
We thank D.Antonov and F. Caristo for a careful reading of the draft and for 
many useful suggestions. We warmly thank M.  Billo',  F. Gliozzi, M. Hasenbusch, A. Nada,  M.Panero and D. Vadacchino for a longlasting fruitful collaboration on the topics discussed in this review.  
\vskip 1cm

\appendix
\section{Appendix:  Useful Formulae}
\label{AppendixA}

Here are some properties of the modular functions which appear in the text. To simplify notations we shall denote $\tilde \tau\equiv -\frac{1}{\tau}$ in the following

The relation with the variables used in the text is:
\begin{align}
&\tau\equiv i\frac{N_t}{2R}~,& q\equiv e^{2\pi i \tau}=e^{-\frac{\pi N_t}{R}} ~, \notag \\
&\tilde\tau \equiv -\frac{1}{\tau}=i\frac{2R}{N_t}~,&\tilde{q}\equiv e^{2\pi i \tilde\tau}=e^{-\frac{4 \pi R}{N_t}} ~.
\end{align}
The Dedekind-$\eta$-function is
\begin{align}
\eta(q)\equiv q^{\frac{1}{24}}\prod_{n=1}^{\infty}(1-q^n)~.
\end{align}
The Eisenstein functions are defined as:
\begin{align}\label{series1}
E_{2k}(q)&\equiv  1+\frac{2}{\zeta(1-2k)}\sumn \frac{n^{2k-1}q^n}{1-q^n}~,\notag\\
\end{align}
where $\zeta(s)$ denotes the Riemann $\zeta$ function defined as follows:

\begin{align}
\zeta(s)\equiv \sumn n^{-s}~,
\end{align}

The Eisenstein functions can be expanded as follows:
\begin{align}
&E_2(q)=1-24q-3\cdot 24 q^2-4\cdot 24 q^3-7\cdot 24 q^4-\cdots\notag \\
&E_4(q)=1+10\cdot 24q+90\cdot 24 q^2+\cdots \notag \\
\end{align}

These functions transform as follows under the modular transformation $\tau\rightarrow -\frac{1}{\tau}$ (notice the inhomogeneous term in the $E_2$ function):

\begin{align}\label{Modular Transformations}
\eta(q)&=(-i\tilde\tau)^{1/2}\eta(\tilde q)=\left(\frac{2R}{N_t}\right)^{\half}\eta(\tilde q)~, \notag \\
E_2(q)&=-\frac{6i}{\pi}\tilde \tau+\tilde \tau^2 E_2(\tilde q)= \frac{12R}{\pi N_t}-\left(\frac{2r}{l}\right)^2 E_2(\tilde q)=\frac{12R}{\pi N_t}\left(1-\frac{\pi R}{3N_t}E_2(\tilde q)\right)~, \notag\\
E_4(q)&=\tilde \tau^4 E_4(\tilde q)=\left(\frac{2R}{N_t}\right)^4 E_4(\tilde q)~.
\end{align}


\providecommand{\href}[2]{#2}\begingroup\raggedright\endgroup
\end{document}